\documentclass[11pt]{article} 
\usepackage{graphicx}
\usepackage{verbatim}
\usepackage{amssymb} 
\usepackage{enumitem}
\usepackage{tabularx}
\usepackage{natbib} 
\usepackage{color, colortbl}
\definecolor{Gray}{gray}{0.9}
\definecolor{pearl}{rgb}{0.94, 0.92, 0.84}
\definecolor{darkgreen}{rgb}{0.0, 0.2, 0.13}
\definecolor{amber}{rgb}{1.0, 0.49, 0.0}
\definecolor{or}{rgb}{0.99, 0.84, 0.69}
\usepackage{amsmath}
\usepackage{graphicx}
\usepackage[colorinlistoftodos]{todonotes}
\usepackage[colorlinks=true, allcolors=blue]{hyperref}
\usepackage{caption}
\usepackage[flushleft]{threeparttable}
\usepackage{siunitx}
\usepackage{float}

\begin{document}
\pagestyle{plain} 
\pagenumbering{arabic}

\begin{center}
ngVLA Memo No. 108\\
\vspace {2mm}
{\Large\bf Feasibility of Self-cal for ngVLA Dynamic Range Requirements} \\

\end{center}

\noindent{\centerline{T. K. Sridharan \& Sanjay Bhatnagar}}

\vspace{5mm}


\section{Background and Motivation} 

The ngVLA science goals require the ability to detect weak emission in the presence of strong peak emission at levels that demand high dynamic ranges in the constructed images. The emission occurring within an observing field in general and their peaks can be exploited to meet the specified dynamic range requirements through self-calibration. When sufficient signal to noise ratio is realized in an observation, self-calibration is almost always used to improve image quality. The feasibility of this approach can be assessed by looking at source counts of background sources serendipitously occurring within the fields of view for each band in relation to achievable noise levels, even in the absence of strong emission from the science target.

An alternative approach is to recognize the broader nature of the question - requiring a certain dynamic range, $DR$, at a targeted science noise level  $\sigma_{science}$ presupposes the presence of bright emission in the field at a corresponding level of detected interferometric flux of $\sim DR \times\sigma_{science}$, by definition. In fact, it is emission from such a background source that drives the most stringent dynamic range requirement for ngVLA observations in the first place. The 45 dB requirement in Band 2 (SCI0113, SYS6103) results from the key science goal targeting deep-field continuum studies of MW-like galaxies at 8 GHz (KSG03), in the presence of a brighter background galaxy source in the field. Would that source allow self-cal? With the problem posed broadly in this way, one can derive a general limit without recourse to background source counts and the attendant Poisson fluctuations of their occurrence, and which depends only on the number of antennas in the array and the science noise level of the observation, independent of the observing band, primary beam size and antenna SEFDs. In the simple case of a point source, this implies a limit on the solution interval at which self-cal can be carried out. This limit is derived below. 

The approach is not fundamentally new, known to many in one form or another and is captured in the maxim roughly stated as  ``{\it ...if you are DR limited, you can selfcal and selfcal will help; if you benefit from selfcal you are (were) DR limited; conversely, if there is not enough SNR for selfcal, you have reached the DR limit your data allows...}'' ({\it e.g.} Bhatnagar, private communication). This synthesis imaging lore and its implications are placed on a formal footing here.

\section{Solution Interval Limit} 

\noindent With \\ \\
$DR =$ dynamic range demanded \\
 $N_A=$ Number of antennas \\
$SNR_{A\_selfcal\_threshold}= $ required SNR per antenna, including all baselines to it, to allow self-cal \\
$t_{sol} =$ self-cal solution interval lower limit (for a given $DR, SNR_{A\_selfcal\_threshold}$ and $N_A$) \\
$t_{int} =$ integration time to reach a science rms of $\sigma_{science}$ \\ \\ 
needing to fulfill a demanded dynamic range $DR$ implies a peak flux in the field of \\
\begin{equation}
S_p = DR\times \sigma_{science} = DR \times SEFD \times \sqrt{2}/\sqrt{N_A(N_A-1)\Delta\nu t_{int}}
\end{equation}

\noindent Similarly, to achieve a SNR of $SNR_{A\_selfcal\_threshold}$ for one antenna for self-cal to be feasible on a solution interval time scale $t_{sol}$, using all baselines to it, the peak flux needed is

\begin{equation}
S_p =  SNR_{A\_selfcal\_threshold} \times SEFD/\sqrt{(N_A-1)\Delta\nu t_{sol}}
\end{equation}

\noindent combining (1) and (2) \\
\begin{equation}
t_{sol}/t_{int} = N_A \times (SNR_{A\_selfcal\_threshold}/DR)^2/2
\end{equation}

\noindent or equivalently,
\begin{equation}
t_{sol} = t_{int}\times N_A \times (SNR_{A\_selfcal\_threshold}/DR)^2/2
\end{equation}

\noindent $t_{sol}$ is the lower limit to the solution interval above which a SNR of $SNR_{A\_selfcal\_threshold}$ for self-cal to be feasible, is achieved. 

While the discussion above considers point sources and peak fluxes as these cases are more readily appreciated, we note that the arguments are equally applicable and agnostic to other source structures present. What matters for the signal-to-noise ratio is the interferometric flux actually detected, irrespective of the structures from which it arises.
A measure of the detected flux in a general image is the quadrature sum of the emission - as adopted in the ngVLA requirements definition.

\begin{table}[h!]
\label{Tab:PhaseFluctuations}
  \begin{threeparttable}
  \caption{Expected SelfCal solution intervals ($N_A=214$; $SNR_{A\_SelfCal\_threshold}=3; SNR_{map} = 31$).}
  \begin{tabular}{|l|l|l|l|l|l|l|l|l|}
    \hline
    Band & Freq. & DR & $t_{sol}/t_{int}$ & $t_{sol1hr}$ &$t_{sol10hr}$  &$t_{sol16hr}$ & $t_{sol100hr}$ & Driving SCI requirement \\
         & (GHz) &(dB)&                  &   (sec)       &   (sec)       &    (sec)      &    (sec)       &   case, integration time \& rms                       \\
    \hline
    1    & 1.2 & 45 & $9.6\times10^{-7}$&   0.003      &    0.03    &    0.06     &  0.35          &\\
    \hline
    \rowcolor{or}
    2    & 8   & 45 & $9.6\times10^{-7}$&    0.003     &    0.03    &  {\bf 0.06} &  0.35          &SCI-0113, 16hr, 0.035$\mu Jy$ \\
    \hline
    3    & 18  & 40 & $9.6\times10^{-6}$&    0.035     &    0.35    &   0.6      &   3.5         &\\ 
    \hline
    \rowcolor{or}
    4    & 27  & 35 & $9.6\times10^{-5}$& {\bf 0.35}    &    3.5     &   6     &   35       &SCI-0113, 1hr, 0.18$\mu Jy$\\
    \hline
    5    & 50  & 32 & $3.8\times10^{-4}$&   1.4        &   14    &   22      &   140       &\\
    \hline
    6    & 70  & 30 & $9.6\times10^{-4}$&   3.5        &   35     &   56     &   350       &\\
    \hline
    6    & 116 & 28 & $2.4\times10^{-3}$&   8.6       &   86   &   138     &   860      &\\
    \hline
  \end{tabular}
  \begin{tablenotes}
      \tiny
    \item (1) The two frequencies with relevant science requirements derived from key science goals are highlighted, vis., 8GHz: SCI-0113, KSG3-011, 16 hr integration, 0.035$\mu Jy$ and 27 GHz: SCI-0113, KSG3-008, 1 hr integration 0.18 $\mu Jy$.
    \item (2)The applicable $t_{sol}$ limit for these cases are shown in bold.
    \end{tablenotes}
  \end{threeparttable}
\end{table}

Table 1 lists $t_{sol}/t_{int}$  and $t_{sol}$ for $SNR_{A\_selfcal\_threshold} = 3$, for different frequencies for which dynamic range requirements are specified in the ngVLA system requirements and for different values of integration times. The choice of $SNR_{A\_selfcal\_threshold} = 3$ is equivalent to 
full map SNRs of 22 and 31 for 107 and 214 antennas for an integration time of $t_{sol}$. The listed science integration times (or equivalently, science noise levels) of 1, 10 and 100 hrs, span the range of integration times in the relevant science goals which drive the DR specifications. The science requirements, and therefore the system requirements, only specify $DR$ for 8 GHz and 27 GHz. Thus, driving $DR$ requirements exist only for these frequencies. For other bands, $DR$ values have been inter-/extrapolated.

As an aside, we note that it is formally incomplete to specify a dynamic range without a targeted science noise level (or integration time) – it matters where the specified dynamic range is placed in the range of observed flux levels. This has strong implications for applicable calibration strategies, in particular, if self-cal would be feasible or not. The science requirements may be revised to state the dynamic range requirements along with the noise levels required.

As can be seen from the table, as the required dynamic range increases, the feasible solution interval decreases. With a higher demanded dynamic range, the peak emission that demands this dynamic range is higher, allowing a smaller solution interval.  For the two frequencies, 8 \& 27 GHz, where relevant science requirements derived from key science goals drive the highest DRs, the minimum solution interval limits indicated are $<1$s, above which $SNR_{A\_selfcal\_threshold} =$ 3 is achieved, for the noise levels (or equivalently, the integration times) called for by the respective KSGs. In other words, there is sufficient SNR to deliver self-cal on the time scales as short as indicated. This is made possible by the high sensitivity of the ngVLA arising from its large collecting area. Thus, very fine-grained tracking of antenna gains is possible through self-cal. Based on past experience, self-cal is the only time-tested path to achieve high dynamic range imaging 
and should be a key calibration strategy for the ngVLA.  The numbers derived show that this is also feasible: for the time scales implied – as short as 0.06 s and 0.35 s at 8 \& 27 GHz – experience indicates a comfortable situation for attaining high dynamic range imaging. While the most sensitive observations will use the most number of antennas, we note that with only 107 antennas considered, the solution interval time scales are still very short, at 0.12 s and 0.7 s at 8 \& 27 GHz (note, however, that the required science integration times also increase). To be more definitive, these solution intervals should be compared with time scales on which gain fluctuations occur. Instrumental fluctuations are clearly on much longer time scales, leaving only fluctuations of ionospheric and tropospheric origins to consider, which we now turn to. 

\section{Atmospheric Phase Fluctuations}

Our approach is to estimate the residual phase fluctuations expected after calibration on various time scales suggested in Table 1 and assess if they would be sufficiently small.  In addressing phase fluctuations we make the reasonable assumption that the VLA site is representative of the ngVLA sites. Carilli \& Holdaway (1997) and Carilli et al (1999) present a framework and analysis for understanding phase fluctuations in general and in the context of measurements using the Atmospheric Phase Interferometer (API) at the VLA site. Importantly, their work predicts expected phase fluctuations remaining after calibration. Butler \& Desai (1999) apply this approach to characterize year round conditions using the API data. While more details can be found in the cited literature, briefly, for fast switching calibration with a calibrator $\theta$ radian away, a cycling time of $t_{cyc}$ and winds aloft of $V_a$, the phase fluctuations are {\it stopped} at an effective baseline scale $b_{eff}$ given by

\begin{equation}
b_{eff} = V_a \times t_{cyc} / 2 + \theta \times h
\end{equation}

\noindent where $h$ is the height at which the dominant phase fluctuations occur. For self-cal, the second term drops out as $\theta \approx 0$ (calibrator is the target, in the primary beam), leading to

\begin{equation}
b_{eff, selfcal} = V_a \times t_{sol} / 2 
\end{equation}

\noindent For the short solution intervals indicated above (0.06 \& 0.35 s; Table 1), and  a nominal $V_a$ of $\sim 10$ m/s (Carilli et al 1999), $b_{eff}$ is $\sim 1-5$ m for 8 and 27 GHz. In other words, the implied residual phase fluctuations of atmospheric origin after self-calibration would be that of an array of only $\sim$ 5-10 m extent. While the strict applicability of the approach to such small scales may be questioned, it is clear that the expected residual fluctuations will be very small.
\begin{figure}[h!]
\begin{center}   
\includegraphics[width=0.73\linewidth]{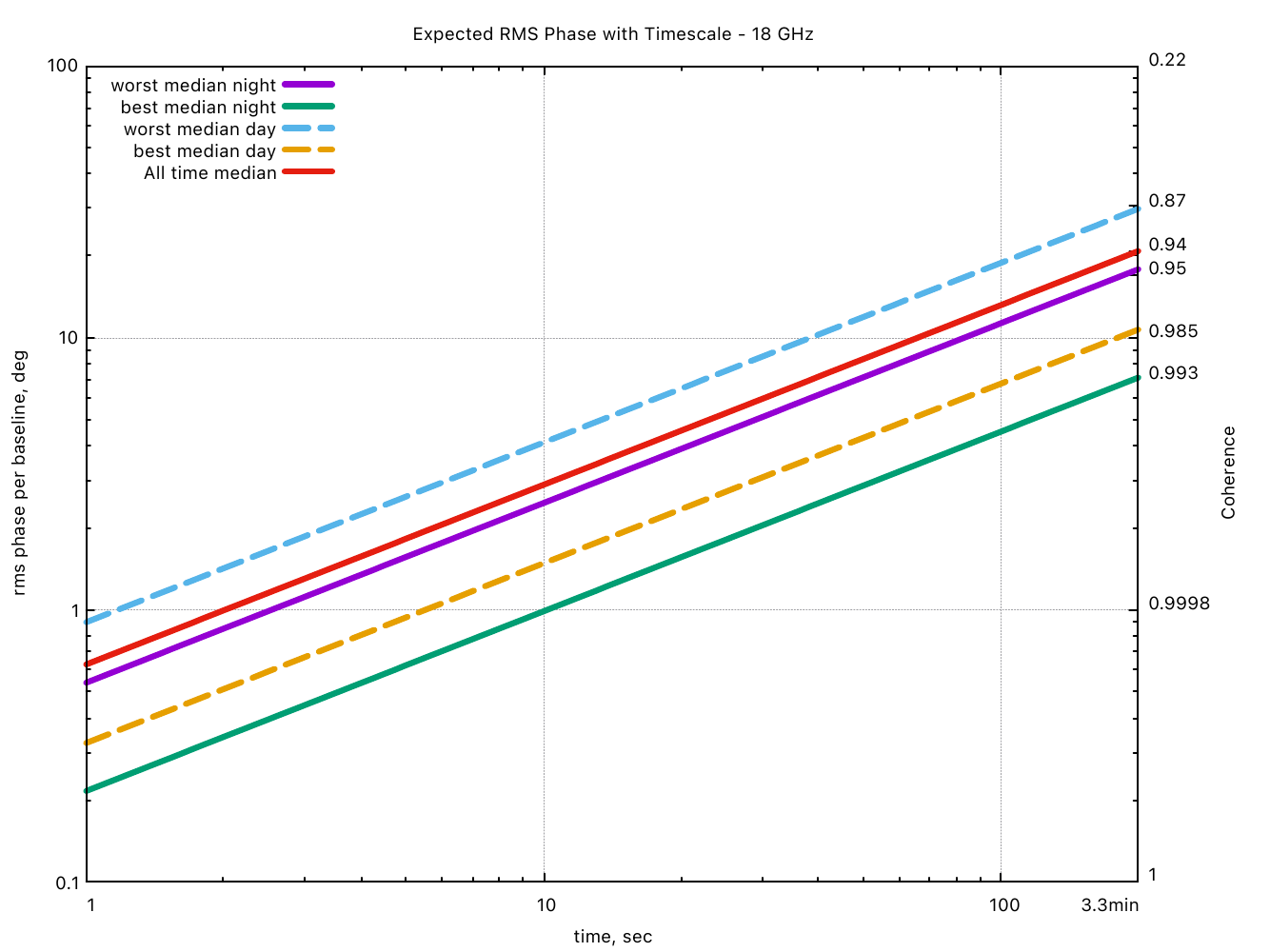}
\vspace{-0.3cm}
\caption{\small Expected residual phase fluctuations after calibration (y-axis) for various time scales (x-axis), derived from one year of data from the Atmospheric Phase Interferometer (based on analysis in Butler \& Desai, 1999). The right y-axis shows the expected coherence ($e^{-\phi^2/2}$) at 18 GHz.}
\vspace{-5mm}
\label{Fig:PhaseFluctuations}
\end{center}
\end{figure}

Figure 1 shows the expected phase fluctuations after calibration as a function of timescale, based on data from API over one year studied by Butler \& Desai (1999). This study constructed and obtained fits to the log structure function (log temporal phase structure function vs log lag) for the API data, using it to predict monthly day and night time medians for the residual phase fluctuations. The results shown in Figure 1 are scaled for a frequency of 18 GHz, chosen to lie between 8 and 27 GHz, the two frequencies with the most stringent driving ngVLA DR requirements. Median values for months showing the worst daytime to the best nighttime conditions are plotted along with the all-time median. As the API operates at 11.3 GHz, the measurements include contributions from both ionospheric and tropospheric components and scaling to higher frequencies as done here therefore results in overestimates, leading to a conservative assessment. 

\section{8 GHz}

Not surprisingly, Figure 1 shows that the residual phase fluctuations at 18 GHz after calibration, even for 1 s solution interval are small, $<$ 1 deg. It is clear that at 8 GHz, the combination of ionospheric and tropospheric fluctuations after calibration (self) will contribute very little to the residual errors. With a background point source driving the DR requirement leading to similar fluxes on all baselines and the solution interval time scale for 8 GHz being orders of magnitude lower than needed, a large margin for the feasibility of selfcal is realized. It is evident from the above discussion and Table 1 that selfcal should be the baseline calibration strategy for band 2 (8 GHz).  

\section{27 GHz}

Moving to 27 GHz, the situation is not as clear. Here, the DR requirement arises from emission in the science target. The possibility that the emission in the beam may be resolved out on some baselines needs to be considered.
There are multiple mitigating factors, however: \\
\noindent (1) As previously noted, the signal to noise ratio and the dynamic range depend on the actual detected fluxes, a measure of which is the quadrature sum of the emission, including distributed emission.   \\
\noindent (2) The baselines on which flux is (partially) resolved out also contribute proportionally smaller image errors, impacting the DR less, by the same factor. The following points and approximate estimates help build some intuition for the impact. In the extreme case, a baseline with no flux does not contribute an image error irrespective of the gain errors present.  If $\sim$ half the antennas have baselines with zero flux, the DR is lowered only by $\sim$ 1.5 dB ({\it i.e.} $\sqrt{N}$; Sridharan et al 2022), although a different total integration time will be required to reach the targeted sensitivity. In general, if a fraction {\it f} of the antennas has baselines with lower fluxes by a factor {\it r} due to resolved source structure, the resulting DR decreases by: 
\begin{equation}
\Delta DR \approx  ( 1- f(1-r))/\sqrt{1-f(1-r^2)}
\end{equation}

\noindent for natural weighting (see Appendix A for a derivation). For baselines to half the number of antennas losing half the flux, $f = 0.5$ and $r = 0.5$ and DR decreases by 0.2 dB. \\
\noindent (3) Source counts imply a $\sim$ 25 $\mu$Jy background source in the beam. \\

We now approach the requirement from a different perspective. For a SNR of 3, the expected antenna based phase error on the self-cal solutions is $\sim$ 20 deg. For the residual phase fluctuations to contribute $<$ 10\% of additional error (equivalent to worsening of the DR by the same factor or 0.4 dB), they are required to be limited to $< 0.45 \times 20 = 9$ deg, or 12 deg for a baseline. From Fig 1, this level of residual phase fluctuation results at $t_{cyc}$ time scale of $\sim$  50 s (for 8 deg, when scaled to 27 GHz), under all time median conditions ($>$ 200 sec under best night time median conditions). This provides 2-3 orders of magnitude additional room, over the 0.35 s achievable $t_{sol}$ (Table 1). For a 50 s solution interval, the required source flux to reach SNR = 3 is $\sqrt{0.35/50}$  $\times$ that required at 0.35 s = 48 uJy, lower by a factor of $\sim$ 12 (24 under best night time median conditions). In other words, even if the science source flux at 27 GHz that demands the 35 dB DR ( 560 $\mu$Jy = $DR \times \sigma =$ 3160 $\times$ 0.18 $\mu$Jy) is resolved out by a factor of $\sim$ 10-20 on some baselines, there is sufficient SNR to meet the threshold for selfcal. In addition, a serendipitous background source expected within the field further alleviates the situation. From source counts, the brightest source within one FWHM at 27 GHz is expected to have a flux of  $\sim$ 25 $\mu$Jy (interpolating using fits in Murphy \& Chary 2017). Including this source for the selfcal SNR requirement, the required flux from the science target is reduced to 40 $\mu$Jy allowing a factor of 15 for loss of flux due to source structure being resolved ($ \sim 30$ for best night time median conditions). While this can be checked against models of source structure, this clearly provides a very comfortable margin. Even with the loss of baselines to some antennas, the DR impact is minimal, as previously outlined. Considering background sources fainter than the brightest also adds further room - it turns out that $\sim$ twice as many background sources are expected at half the flux (Murphy \& Chary, 2017). 

The best performance requirements of the ngVLA should not be demanded at all times, and seeking to fulfill them even under all-time median conditions is quite aggressive. Considering all-time median conditions (Fig. 1, red line) from one year of API data (Butler \& Desai, 1999), one can see that on 3 minutes time scales, a coherence loss factor of 0.94 is predicted at 18 GHz after calibration. On these time scales, the current baseline antenna drive requirements allow fast switching to a target $3^\circ$ away with 90\% observing efficiency (SYS 1061: 90\% calibration efficiency goal). This forms the back-up calibration strategy. Further, the solution intervals at which self-cal is feasible are very small compared to the fluctuation time scales (Table 1 \& Fig 1). Therefore, both fast switching (needed for good  initial models for self-cal) and self-cal are feasible. Employing self-cal would allow the switching time scale to be extended beyond 3 minutes, delivering observing efficiencies well above 90\% and the attainment of high dynamic range imaging. 

In summary, selfcal is a feasible strategy for bands 2 \& 4 (8 \& 27 GHz) to mitigate phase fluctuations on relevant time scales.  

\section{Achievable Dynamic Range}

The next question is what dynamic range can be achieved through self-calibration. As we show below, by definition, this is self-consistently equal to the required dynamic range, $DR$, which we started with, as long as self-cal is feasible on the required time scales. 

The per antenna phase error on the self-cal solutions for the above context ({\it i.e.} the combination  $ SNR_{A\_selfcal\_threshold},DR, t_{sol},t_{int}$ and $\sigma_{science}$), that can be ideally achieved,  is  $\sim 1/SNR_{A\_selfcal\_threshold}$  radian. 

\begin{equation}
\phi_{\sigma,A,selfcal,t_{sol}} =  1/SNR_{A\_selfcal\_threshold}
\end{equation}

\noindent using (3), we have

\begin{equation}
\phi_{\sigma,A,selfcal,t_{sol}} =   (N_A \times t_{int}/t_{sol})^{1/2}/DR
\end{equation}

\noindent This would be the residual phase error on each antenna after applying self-cal solutions. The SNR on a map made with baselines to one antenna with these self-cal corrections is, then,
\begin{equation}
SNR_{map,1Antenna} =  1/\phi_{\sigma,A,selfcal,t_{sol}} =  DR /(N_A \times t_{int}/t_{sol})^{1/2}
\end{equation}

\noindent The SNR for the full map, combining all antennas and for the full integration time would be higher by  $(N_A \times t_{int}/t_{sol})^{1/2}$. Therefore, from (10), we have,

\begin{equation}
SNR_{map,selfcal,t_{int}} = DR 
\end{equation}

\noindent This SNR is the map dynamic range achieved. Thus,

\begin{equation}
DR_{map,selfcal,t_{int}} = DR
\end{equation}
In short, we reach a general, self-consistent conclusion that as long as the solution interval time scale on which self-cal is feasible is smaller than the time scales of corrupting residual fluctuations, the required dynamic range is achievable.

\section{Other Prospects}

If self-cal is feasible on smaller time scales, as is the case here, other benefits accrue. The correspondence between time and bandwidth can be exploited to split the spectral coverage into sub-bands – it is the $t_{sol} \times \Delta\nu_{sol}$ product, the number of independent data samples, that matters - to handle changing source structure over the band, which would otherwise impact self-cal. Parameterization of the spectral variation may be a more efficient approach. Given the additional room, other wide bandwidth and direction dependent effects would also become amenable to elimination through self-cal. The fact that these effects are only expected to corrupt the observed visibilities on longer time scales and can therefore tolerate longer solution intervals, further eases the situation. Effectively, the high sensitivity of the ngVLA holds out the possibility of reaching thermal noise limited dynamic range, at least in some bands and in some cases.

\section{Conclusion}

In conclusion, selfcal is feasible for achieving the high image dynamic ranges required to satisfy ngVLA science goals in bands 2 \& 4 and should be the preferred strategy. Bracketed by these two bands, the conclusion is also applicable to an interpolated DR requirement for band 3, although a specific key science goal based requirement does not exist.  An important outcome of the above analysis is that, water vapor radiometry, the baseline strategy currently envisaged for these bands, is no longer required (and in fact may be undesirable for the lower bands), as the image dynamic range requirements for bands 2, 3 \& 4 can be delivered through selfcal. However, as the band 4 receiver covers the water line, it provides water vapour radiometry as an additional input to correct for phase errors. This also opens up the possibility of using Band 4 water vapor radiometry for Bands 5 \& 6. The limits from extension of the selfcal approach to higher frequencies remain to be worked out.


\newpage

\appendix
\section{Partially Resolved Sources}

We derive an approximate expression for the expected dynamic range when a source is partially resolved. Considering a fraction $f$ of the antennas having baselines suffering a flux reduction by a factor $r$, 

\begin{equation}
S \approx N(1-f) + Nrf = N[1-f(1-r)]
\end{equation}
\begin{equation}
\sigma \approx \sqrt{N(1-f)\epsilon^2 + Nfr^2\epsilon^2} = \sqrt{N}\epsilon\sqrt{1-f(1-r^2)}
\end{equation}
\noindent where $S$ is the signal and $\epsilon$ is the fractional error attributable to the set of baselines to one antenna. 
\begin{equation}
DR = S/\sigma \approx \frac{\sqrt{N}}{\epsilon} \frac{1-f(1-r)}{\sqrt{1-f(1-r^2)}}
\end{equation}
\noindent The nominal dynamic range expected for $N$ antennas is $\sqrt{N}/\epsilon$ with no loss of flux (see also Sridharan et al 2022). Therefore, the reduction in the dynamic range can be written as 

\begin{equation}
\Delta DR \approx \frac{1-f(1-r)}{\sqrt{1-f(1-r^2)}}
\end{equation}
  

\end{document}